\documentclass{piparticle-final}
\usepackage[utf8]{inputenc}
\usepackage{amsmath}
\usepackage{amssymb}
\usepackage{graphicx}

\usepackage{cite} 
\usepackage{epstopdf}

\begin{document}

\volume{7}               
\articlenumber{070015}   
\journalyear{2015}       
\editor{C. A. Condat, G. J. Sibona}   
\received{20 November 2014}     
\accepted{15 September 2015}   
\runningauthor{I. C. Ramos \itshape{et al.}}  
\doi{070015}         

\title{Adapting a Fourier pseudospectral method to Dirichlet boundary conditions for Rayleigh--B\'enard convection}

\author{I. C. Ramos,\cite{inst1} \hspace{0.2em} 
            C. B. Briozzo\cite{inst1}\thanks{E-mail: briozzo@famaf.unc.edu.ar}
 }

\pipabstract{
We present the adaptation to non--free boundary conditions of a pseudospectral method based on the (complex) Fourier transform. The method is applied to the numerical integration of the Oberbeck--Boussinesq equations in a Rayleigh--B\'enard cell with no-slip boundary conditions for velocity and Dirichlet boundary conditions for temperature. We show the first results of a 2D numerical simulation of dry air convection at high Rayleigh number ($R\sim10^9$). These results are the basis for the later study, by the same method, of wet convection in a solar still.
}

\maketitle

\blfootnote{
\begin{theaffiliation}{99}
   \institution{inst1} Facultad de Matem\'atica, Astronom\'ia y F\'isica, Universidad Nacional de C\'ordoba, X5000HUA C\'ordoba, Argentina.
\end{theaffiliation}
}

\section{Introduction}

Experimental observations \cite{DePaul2009} show that the onset of a turbulent convective flux can significantly enhance the efficiency of a basin-type solar still, but until now a theoretical explanation is lacking. Any adequate hydrodynamical simulation must incorporate the effects of moisture and condensation. Recent works \cite{Pauluis2008,Pauluis2010} show that this can be achieved through a Boussinesq-like approximation, which simplifies considerably the problem. However, realistic simulations are still demanding, given the need to resolve fine flux details and cope with Rayleigh numbers up to $\sim 10^9$ \cite{Weidauer2012}.

Spectral methods \cite{Fornberg1999} are well suited for this kind of tasks, and have many attractive features: they are simple to implement, show much better resolution and accuracy properties than finite difference or finite volume methods, and are highly efficient in large-scale simulations \cite{Mercader2009}. Fourier-based pseudospectral methods are the simplest and fastest, since the discretized spatial differential operators are local, nonlinear terms can be computed through Fast Fourier Transform (FFT) convolutions, and solving the Poisson equation originating from the incompressibility (divergence-free) condition is almost trivial. Nevertheless, they usually work only for free (in fact, periodic) boundary conditions (BCs).

The presence of non-free ({\it e.g.}, Dirichlet or Neumann) BCs introduces additional complications. For example, two-dimensional Rayleigh-B\'enard convection with laterally periodic BCs can be treated by using a spectral Galerkin--Fourier technique in the horizontal coordinate and a collocation-Chebyshev method in the vertical one \cite{Mercader2006}, but vertical derivatives must then be computed by matrix multiplication. On a grid with $N$ horizontal and $M$ vertical points, this needs the solution of a linear system of dimension $M$ for each of the $N$ horizontal Fourier modes, at each time-integration step.

Another complication with non-free BCs arises from the need to fulfill numerically the divergence-free condition, leading mainly to two different groups of methods (see Ref. \cite{Mercader2009} and references therein). In a first group the velocity field is written in terms of scalar potentials such that the divergence-free condition is satisfied by construction, {\it e.g.}, in the 2D streamfunction-vorticity formulation or the 3D decomposition into toroidal and poloidal velocity potentials. In these methods, pressure is not present in the equations, but they lead to systems of higher-order partial differential equations with coupled BCs. In a second group, a primitive variable formulation of the equations is adopted and projection methods \cite{Chorin1968} are used to decouple velocity and pressure. These methods use a specific splitting of the equation system based on the chosen time-integration scheme, and determine pressure by projecting an appropriate velocity field onto a 
divergence-free space, 
leading to predictor-corrector algorithms. Besides the problem of correctly specifying the pressure BCs \cite{Gresho1987,Sani2006}, these methods require solving a Poisson equation for the pressure at each time-integration step. On a ${\cal N}=N\times M$ grid, the best Fourier-based Fast Poisson Solvers (FPS) have operation counts ${\cal O}({\cal N}\log_2{\cal N})$ for the lowest (second) order discretization (and significantly worse for higher orders) \cite{Fuka2015,Braverman1998}, and those using GMRES are $K{\cal O}({\cal N})$ with $K \gtrsim 100$ \cite{Gresho2008,Gresho2010,Saad1986}.

In this work, we will show how a Fourier-based pseudospectral method can be adapted to simple non-free (but periodic) BCs without losing its more appealing features. This is a first step towards building a pseudospectral simulation of wet air convection inside a basin-type solar still, and must be considered just as a proof of concept.

\section{System}

We consider 2D dry air convection in a Rayleigh--B\'enard cell of width $L=1\,$m and height $H=0.5\,$m, close to room temperature and with temperature differences $\Delta T=T_h-T_c$ up to $\sim 65$K between the hot lower ($T=T_h$) and cold upper ($T=T_c$) plates (roughly the parameters of a real still \cite{DePaul2009}). Discarding thermal fluctuations and the heat generated by viscous dissipation, and assuming an incompressible fluid across which all thermodynamical parameters change little, the dynamics is given by \cite{Landau1966}

\begin{align}
\label{eq:momentum_transp_sim}
\rho (\partial_{t}+\mathbf{u}\cdot\mathbf{\nabla}) \mathbf{u}
&=-\mathbf{\nabla }P+\eta \nabla ^{2}\mathbf{u}-\rho g\hat{\mathbf{z}}, \\
\label{eq:heat_transp_sim}
\rho c_{p} (\partial_{t}+\mathbf{u}\cdot\mathbf{\nabla}) T
&=K\nabla ^{2}T ,\\
\label{eq:mass_transp_sim}
\mathbf{\nabla }\cdot \mathbf{u} &=0.
\end{align}
In these equations, the dynamical variables are the density $\rho $, the velocity $\mathbf{u}$, the temperature $T$, and the pressure $P$. The parameters are the shear (or dynamic) viscosity $\eta $, the constant pressure specific heat capacity $c_{p}$, the thermal conductivity $K$, and the gravitatioinal acceleration $g$.

The Boussinesq approximation \cite{Cross1993a} consists in discarding the  dependence of $\eta $, $c_{p}$, and $K$ on temperature and density, keeping 

\begin{equation}
\rho =\bar{\rho}[ 1-\alpha (T-\bar{T}) ] 
\end{equation}
in the buoyancy term ($-\rho g \hat{\mathbf{z}}$ in Eq.~\ref{eq:momentum_transp_sim}) but otherwise setting $\rho=\bar{\rho}$ elsewhere. Here, $\bar{T}=\frac12(T_h+T_c)$ is a reference temperature, $\bar{\rho}$ is a reference density (that of air at normal temperature and pressure), and $\alpha$ is the thermal expansion coefficient. Dropping the bars signifying reference quantities, absorbing some constants into the pressure gradient term, and defining the viscous diffusivity (or kinematic viscosity) $\nu =\eta/\rho$ and the thermal diffusivity $\kappa =K/(\rho c_p)$, we obtain the (dimensional) Oberbeck--Boussinesq equations \cite{Cross1993a}

\begin{align}
\label{eq:WOB_dim_momentum}
(\partial_{t}+\mathbf{u}\cdot \mathbf{\nabla }) \mathbf{u}
&=-\mathbf{\nabla } (P/\rho) +\nu \nabla ^{2}\mathbf{u}
-\alpha gT\hat{\mathbf{z}}, \\
\label{eq:WOB_dim_heat}
(\partial_{t}+\mathbf{u}\cdot\mathbf{\nabla}) T
&=\kappa \nabla ^{2}T ,\\
\label{eq:WOB_dim_mass}
\mathbf{\nabla }\cdot \mathbf{u} &=0.
\end{align}
Assuming perfect thermal contact with the lower ($z=0$) and upper ($z=H$) plates, these equations admit the stationary conductive solution

\begin{align}
\label{eq:Modelo_rect_sscp}
\mathbf{u} &=0, & \notag\\
T_0(z) &= \bar{T}-\frac{\Delta T}{H} z', & \notag\\
P_0(z) &= \bar{P}+\rho g\alpha\left( 
         -\bar{T}z'+\frac{\Delta T}{2H} z'^2 \right),
\end{align}
where $\bar{P}$ is a reference ({\it e.g.} normal atmospheric) pressure, and $z'=z-H/2$.

We now scale lengths with the cell height $H$, times with the characteristic vertical thermal diffusion time $t_c=H^2/\kappa$, and temperatures with the temperature difference $\Delta T$. The nondimensional lengths, times, and velocities are then

\begin{align}
\label{eq:WOB_rescaling_fund}
\mathbf{r}' &=\frac{1}{H}\mathbf{r} , &
t' &=\frac{\kappa}{H^{2}}t , &
\mathbf{u}=\frac{\kappa}{H}\mathbf{u}' . &
\end{align}
We also define the nondimensional temperature $\theta$ and pressure $P'$ by

\begin{align}
\theta &= \frac{1}{\Delta T}\left(T-T_0(z)\right) , & \notag\\
P' &= \frac{H^2}{\kappa\nu}\left(P-P_0(z)\right).
\end{align}
Substituting into Eqs. (\ref{eq:WOB_dim_momentum})--(\ref{eq:WOB_dim_mass}), discarding the primes for simplicity, and absorbing all pure gradient terms into the pressure, we get the dimensionless Oberbeck--Boussinesq equations \cite{Cross1993a}

\begin{align}
\label{eq1}
\sigma^{-1}(\partial_t + {\mathbf u}\cdot{\mathbf\nabla}){\mathbf u} &= -{\mathbf\nabla}P + \theta\hat{{\mathbf z}} + \nabla^2{\mathbf u},\\
\label{eq2}
(\partial_t + {\mathbf u}\cdot{\mathbf\nabla})\theta &= Ru_z + \nabla^2\theta,\\
\label{eq3}
{\mathbf\nabla}\cdot{\mathbf u} &= 0,
\end{align}
where

\begin{equation}
R = \frac{g\alpha\Delta T H^3}{\kappa\nu}
\end{equation}
is the Rayleigh number and $\sigma=\nu/\kappa$ is Prandtl's number ($\simeq 0.7$ for dry air). The BCs we adopt are periodic in the horizontal direction, and homogeneous Dirichlet for both velocity ${\mathbf u}$ (no-slip BCs) and temperature $\theta$ (perfect thermal contact) on the lower and upper plates.

Note that Eq. (\ref{eq3}) is not a differential equation but a constitutive relationship, expressing the incompressibility of the flux. In fact, the pressure term in Eq. (\ref{eq1}) is computed by enforcing Eq.(\ref{eq3}), which gives

\begin{equation}
\label{eq4}
\nabla^2 P = -\sigma^{-1}\Sigma_{i,j}\partial_i\partial_j(u_i u_j)
            +\partial_z\theta,
\end{equation}
where $i,j=x,z$. In primitive variable integration schemes, this Poisson equation must be solved with adequate BCs at each time-step \cite{Mercader2009,Gresho1987}, to insure ${\mathbf\nabla}\cdot(\partial_t{\mathbf u})=0$.

\section{Helmholtz decomposition}

Given a vector field~$\mathbf{f}$, twice continuously differentiable, Helmholtz's Theorem \cite{Helmholtz1858} states that it can be decomposed as

\begin{equation}
\label{eq5}
\mathbf{f} = \mathbf{f}_{\|} + \mathbf{f}_{\perp}.
\end{equation}
Here, $\mathbf{f}_{\|}$ and $\mathbf{f}_{\perp}$ are the longitudinal (or irrotational) and the transverse (or solenoidal) components of the field \cite{Chorin1968}, respectively, with

\begin{align}
\label{eq6}
\mathbf{\nabla}\times\mathbf{f}_{\|}&=0, &
\mathbf{\nabla}\cdot\mathbf{f}_{\perp}&=0.
\end{align}
We are now going to rewrite Eq. (\ref{eq1}) in the form

\begin{equation}
\label{eq7}
\partial_t\mathbf{u} = \sigma\left( \mathbf{f} - \mathbf{\nabla}P \right)
\end{equation}
with

\begin{equation}
\label{eq8}
\mathbf{f} = -\sigma^{-1}({\mathbf u}\cdot{\mathbf\nabla}){\mathbf u} + \theta\hat{{\mathbf z}} + \nabla^2{\mathbf u}.
\end{equation}
Then, if $\mathbf{\nabla}\cdot\mathbf{u}=0$ initially, the incompressibility condition Eq. (\ref{eq3}) requires

\begin{equation}
\label{eq9}
\mathbf{\nabla}\cdot\left( \mathbf{f} - \mathbf{\nabla}P \right) = 0.
\end{equation}
This amounts to requiring the field $\mathbf{f} - \mathbf{\nabla}P$ to be purely transverse, that is

\begin{equation}
\label{eq10}
\left( \mathbf{f} - \mathbf{\nabla}P \right)_{\|}
 = \mathbf{f}_{\|} - \mathbf{\nabla}P = 0,
\end{equation}
where we used $\left(\mathbf{\nabla}P \right)_{\|} \equiv \mathbf{\nabla}P$, since the pressure gradient is purely longitudinal. This shows that the \emph{only} effect of the term $\mathbf{\nabla}P$ in Eq. (\ref{eq7}) is to cancel the longitudinal component of $\mathbf{f}$. Equation (\ref{eq7}) can then be set as

\begin{equation}
\label{eq11}
\partial_t\mathbf{u} = \sigma\mathbf{f}_{\perp} ,
\end{equation}
\emph{with no explicit reference to a pressure field}.

In 2D and in free space (that is, disregarding surface terms), the longitudinal and transverse components of $\mathbf{f}$ can be computed as the projections

\begin{align}
\label{eq12}
\tilde{\mathbf{f}}_{\|}
 &= \left(\tilde{\mathbf{f}}\cdot\hat{\mathbf{k}}\right)\hat{\mathbf{k}} ,&
\tilde{\mathbf{f}}_{\perp}
 &= \left(\tilde{\mathbf{f}}\cdot\hat{\mathbf{k}}'\right)\hat{\mathbf{k}}' 
\end{align}
of its Fourier transform~$\tilde{\mathbf{f}}$ along the unit vectors

\begin{align}
\label{eq13}
\hat{\mathbf{k}} &= \frac{\left(k_x,k_z\right)}{\sqrt{k_x^2+k_z^2}}, &
\hat{\mathbf{k}}' &= \frac{\left(-k_z,k_x\right)}{\sqrt{k_x^2+k_z^2}}.
\end{align}
However, on a finite domain, the surface terms cannot be ignored, since they are essential for $\mathbf{f}_{\perp}$ to have the correct BCs, which by Eq. (\ref{eq11}) are the same as those for $\mathbf{u}$ in Eq. (\ref{eq1}). Using Eq. (\ref{eq13}), the field $\tilde{\mathbf{f}}_{\perp}$ in Eq. (\ref{eq12}) can be seen to be a \emph{particular} solution (the free--space solution) of the Poisson equation

\begin{equation}
\label{eq14}
\nabla^2{\mathbf f}_{\perp}
= {\cal F}^{-1}\{ (-k_z,k_x)(-k_z\tilde{f}_x+k_x\tilde{f}_z) \},
\end{equation}
where ${\cal F}^{-1}$ stands for the inverse Fourier transform. To be able to impose the required BCs, we need the \emph{general} solution of this equation, which we can get by adding the general solution of the corresponding \emph{homogeneous} equation $\nabla^2\mathbf{f}_{\perp}=0$.

The required transverse component of $\mathbf{f}$ can then be redefined as

\begin{equation}
\label{eq15}
\mathbf{f}_{\perp} = \mathbf{v} + \mathbf{w},
\end{equation}
with

\begin{equation}
\label{eq16}
\tilde{\mathbf{v}} =  (-k_z,k_x)\frac{-k_z\tilde{f}_x+k_x\tilde{f}_z}{k_x^2+k_z^2}
\end{equation}
and

\begin{align}
\label{eq17}
\nabla^2\mathbf{w}&=0, & \mathbf{\nabla}\cdot\mathbf{w}&=0 ,
\end{align}
where the last equation is needed to insure the transversality of $\mathbf{w}$ and hence of $\mathbf{f}_{\perp}$. This requirement can be automatically fulfilled by writing $\mathbf{w}$ explicitly as

\begin{equation}
\label{eq18}
\tilde{\mathbf{w}} = (-k_z,k_x)\tilde{w}
\end{equation}
with the \emph{scalar} field $w$ satisfying

\begin{equation}
\label{eq19}
\nabla^2 w = c
\end{equation}
where $c$ is a constant. Noting that Eq. (\ref{eq16}) can also be rewritten as

\begin{align}
\label{eq20}
\tilde{\mathbf{v}} &= (-k_z,k_x)\tilde{v} , &
\tilde{v} &= \frac{-k_z\tilde{f}_x+k_x\tilde{f}_z}{k_x^2+k_z^2},
\end{align}
we can also rewrite Eq. (\ref{eq15}) as

\begin{equation}
\label{eq21}
\tilde{\mathbf{f}}_{\perp} = (-k_z,k_x) (\tilde{v} + \tilde{w}),
\end{equation}
which shows explicitly the transversality of $\mathbf{f}_{\perp}$. The determination of the value of $c$, and the treatment of possible divergences in $\tilde{v}$ at $\mathbf{k}=0$, are closely related and will be dealt with in the next section. The BCs for $\mathbf{w}$ at the lower and upper plates can be obtained from those for $\mathbf{f}_{\perp}$, and are $\mathbf{w} = -\mathbf{v}$; the BCs on the horizontal direction are periodic but otherwise free, and will be automatically fulfilled by the constructive procedure for $\mathbf{w}$ given in the next section.

\section{Ultra-fast Laplace solver}

We start by solving Eq. (\ref{eq19}) with $c=0$, that is Laplace's equation, on the rectangular domain $0\leq x\leq L$, $0\leq z\leq H$, which is an elementary problem in harmonic analysis. Over an \emph{unbounded} domain, the solutions have the form $e^{i\lambda x} e^{\lambda z}$, where $\lambda$ is an arbitrary separation constant; the particular solutions for the case $\lambda=0$ are $1$, $x$, $z$, and $xz$. Periodicity in $x$ on $[0,L]$ imposes $\lambda=2\pi p/L$ with $p\in\mathbb{Z}$; the general solution is then

\begin{equation}
\label{eq22}
w = \sum_p c_p e^{-i2\pi px/L}e^{-2\pi pz/L}+ a + bz \, ,
\end{equation}
with $c_p$, $a$, and $b$ (possibly complex) constants. For convenience, and without loss of generality, we will rewrite it in the form

\begin{align}
\label{eq23}
w &= \sum_{p} e^{-i2\pi px/L}\left[ a_{p}\cosh \frac{2\pi pz'}{L} \right.\notag\\
 &\left. + (1-\delta_{p,0}) b_{p}\sinh \frac{2\pi pz'}{L} + \delta_{p,0}b_0 z' \right] ,
\end{align}
where $a_p$ and $b_p$ are constants, and $z'=z-H/2$. Here, we have used that $\cosh(0)=1$ to absorb all constant terms in $a_0$, and used that $\sinh(0)=0$ to absorb the linear term in $z'$ as the particular term of the sum for $p=0$, leaving Eq. (\ref{eq23}) explicitly in the form of a (complex) Fourier series in $x$. The hyperbolic and linear functions in $z'$ can, in turn, be rather trivially expanded as complex Fourier series in $z$ on the interval $[0,H]$, leaving $w$ in the form of the double Fourier series

\begin{align}
\label{eq24}
w &= \notag\\
&\sum_{p,q} e^{-i2\pi px/L} e^{-i2\pi qz/H} \left[ a_{p}\tilde{C}_{pq} + b_{p}\tilde{S}_{pq} \right] ,
\end{align}
where $q\in\mathbb{Z}$ and $\tilde{C}_{pq},\tilde{S}_{pq}$ are the expansion coefficients. Discretizing $w$ on a coordinate grid $(x_n,z_m)=(n\Delta x/N,m\Delta z/M)$ restricts the range of $p$ and $q$ respectively to $[0,N-1]$ and $[0,M-1]$ (so the horizontal and vertical wavenumbers are below the respective Nyquist frequencies), reducing Eq. (\ref{eq24}) to a (double) discrete Fourier transform (DFT), whose coefficients

\begin{equation}
\label{eq25}
\tilde{w}_{pq} = a_{p}\tilde{C}_{pq} + b_{p}\tilde{S}_{pq}
\end{equation}
give the discretization of $w$ on the wavenumber grid. The matrices $\tilde{C}_{pq}$ and $\tilde{S}_{pq}$ are given explicitly by

\small
\begin{align}
\label{eq:Laplace_Cpq_def}
&\tilde{C}_{pq} = \\
&\begin{cases} 
\dfrac{1}{M}
\dfrac{\sinh\left( k_{x,p}\frac{H}{M}\right) \sinh\left(k_{x,p}\frac{H}{2}\right) }
{\cosh\left( k_{x,p}\frac{H}{M}\right) -\cos\left( k_{z,q}\frac{H}{M}\right) }, &
\left( p,q\right) \neq \left( 0,0\right) \\
1, &\left( p,q\right) =\left( 0,0\right)
\end{cases} \notag\\
\label{eq:Laplace_Spq_def}
&\tilde{S}_{pq} = \\
&\begin{cases}
-\dfrac{1}{M}
\dfrac{i\sin\left( k_{z,q}\frac{H}{M}\right) \sinh\left(k_{x,p}\frac{H}{2}\right) }
{\cosh\left( k_{x,p}\frac{H}{M}\right) -\cos\left( k_{z,q}\frac{H}{M}\right) },\ &
p\neq 0 \\ 
-\dfrac{1}{M}
\dfrac{i\sin \left( k_{z,q}\frac{H}{M}\right) \frac{H}{2}}
{1-\cos \left( k_{z,q}\frac{H}{M}\right) }, & p=0, q\neq 0 \\ 
0, & \left( p,q\right) =\left( 0,0\right), \notag
\end{cases}
\end{align}
\normalsize
where $k_{x,p}=2\pi p/L$ and $k_{z,q}=2\pi q/H$. Here we have preferred, for simplicity when writing the numerical code, to replace the intervals $0\leq p\leq N-1$ and $0\leq q\leq M-1$ by the equivalent intervals $-N/2\leq p\leq N/2$ and $-M/2\leq q\leq M/2$, with their respective extreme points identified.

Equation (\ref{eq25}) then provides the general solution of $\nabla^2 w=0$ on $[0,L]\times[0,H]$, discretized on the wavenumber grid. The corresponding general solution of Eq. (\ref{eq19}) can then be expressed formally as

\begin{equation}
\label{eq28}
\tilde{w}_{pq} = a_{p}\tilde{C}_{pq} + b_{p}\tilde{S}_{pq}
               - \frac{c\delta_{p,0}\delta_{q,0}}{k_{x,p}^2+k_{z,q}^2}.
\end{equation}
Then from Eqs. (\ref{eq20}) and (\ref{eq21}), the discretization of $\mathbf{f}_\perp$ on the wavenumber grid will be expressed as

\begin{align}
\label{eq29}
\tilde{\mathbf{f}}_{\perp,pq} &= (-k_{z,q},k_{x,p})
\left[\vphantom{\frac{y_1^2}{y_1^2}} a_{p}\tilde{C}_{pq} + b_{p}\tilde{S}_{pq} \right.\notag\\
 &\left.+ \frac{-c\delta_{p,0}\delta_{q,0}-k_{z,q}\tilde{f}_{x,p}+k_{x,p}\tilde{f}_{z,q}}
        {k_{x,p}^2+k_{z,q}^2} \right].
\end{align}
The constant $c$ can now be formally chosen to cancel the possible divergence at $\mathbf{k}=0$; in practice, we set $c=0$ and redefine $\tilde{v}_{00}$ to be an arbitrary but finite constant, which without loss of generality can also be taken to vanish.

Imposing the BCs $\mathbf{w}=-\mathbf{v}$, or equivalently $\mathbf{f}_\perp=0$, at $z=0$ and $z=H$, is achieved as follows: First we note that any field discretized on the wavenumber grid through a DFT of its discretization on the coordinate grid is automatically periodic in \emph{both} the horizontal and vertical directions, so the BCs at $z=0$ and $z=H$ will give identical sets of equations. Next we use the fact that

\begin{equation}
\label{eq30}
\mathbf{f}_{\perp,n0} = \sum_{p,q} e^{-ik_{x,p}x_n} \tilde{\mathbf{f}}_{\perp,pq} \, ,
\end{equation}
together with the completitude of the DFT, to rewrite the BC at $z=0$ as

\begin{equation}
\label{eq31}
\sum_q \tilde{\mathbf{f}}_{\perp,pq} = 0 \quad \forall p.
\end{equation}
We then use Eq. (\ref{eq29}) and the parity of $\tilde{C}_{pq}$ (even in $q$) and $\tilde{S}_{pq}$ (odd in $q$) to obtain the conditions

\begin{align}
\label{eq32}
a_p \sum_q \tilde{C}_{pq} &= -\sum_q \tilde{v}_{pq} \, , & \notag\\
b_p \sum_q k_{z,q} \tilde{S}_{pq} &= -\sum_q k_{z,q} \tilde{v}_{pq} \, ,
\end{align}
from which the coefficients $a_p$ and $b_p$ can be immediately retrieved. Finally, substituting into Eq. (\ref{eq29}) leads to

\begin{align}
\label{eq33}
\tilde{\mathbf{f}}_{\perp,pq} &= (-k_{z,q},k_{x,p})
\left[ \tilde{v}_{pq} - c_{pq}\sum_q \tilde{v}_{pq} \right. \notag\\
&\left.- s_{pq}\sum_q k_{z,q}\tilde{v}_{pq} \right],
\end{align}
where it is understood that we take $\tilde{v}_{00}=0$, and we have introduced the normalized matrices

\begin{align}
\label{eq34}
c_{pq} &= \frac{\tilde{C}_{pq}}{\sum_{q'}\tilde{C}_{pq'}}, &
s_{pq} &= \frac{\tilde{S}_{pq}}{\sum_{q'}k_{z,q'}\tilde{S}_{pq'}}.
\end{align}

Equations (\ref{eq33}) and (\ref{eq20}) show that the transverse field $\tilde{\mathbf{f}}_{\perp,pq}$ on the wavenumber grid can be obtained directly in terms of the non-transverse field $\tilde{\mathbf{f}}_{pq}$ without needing, at any point, to return to the coordinate grid. Moreover, Eq. (\ref{eq34}) shows that $c_{pq}$ and $s_{pq}$ are given matrices that can be computed just once at the start of the simulation, as is the denominator $k_{x,p}^2+k_{z,q}^2$ in Eq. (\ref{eq20}). The only tasks to be performed at each time-step are then: first, computing the scalars $\tilde{v}_{pq}$ from $\tilde{\mathbf{f}}_{pq}$, requiring three multiplications per grid point; second, computing the sums in Eq. (\ref{eq33}), which requires one multiplication per grid point; third, multiplying them by $c_{pq}$ and $s_{pq}$, costing two multiplications per grid point; and fourth, multiplying by $k_{z,q}$ and $k_{x,p}$ at each grid point. The total cost of 
obtaining $\tilde{\mathbf{f}}_{\perp,pq}$ is then $8{\cal N}$ multiplications on 
a ${\cal N}=N\times M$ grid, thus outperforming even the best FPS by a significant factor on large grids.

It must be noted that the method introduced here has some similarity to the streamfunction--vorticity formulation \cite{Mercader2009}, in the sense that the scalar fields $\tilde{v}$ and $\tilde{w}$ play a role similar to these potentials. However, in our method they are not taken as \emph{dynamical} variables, and the evolution equations are not formulated in terms of them but of primitive variables. The method presented here shows also a strong similarity with projection methods \cite{Chorin1968,Gresho1987}, but differently from them, pressure is not computed along the time evolution and, in fact, does no longer appear in the evolution equations.

\section{Algorithm outline}

We present now an outline of the numeric algorithm as we implemented it in the simulations.

Initialization:
\begin{itemize}
  \item Take zero velocity and Gaussian white noise for the temperature (with amplitude of the order of thermal noise), discretized on the coordinate grid, and take their FFT to get $\tilde{u}_{x,pq}$, $\tilde{u}_{z,pq}$, $\tilde{\theta}_{pq}$ on the wavenumber grid.
  \item Pre-compute (just once) the matrices $c_{pq}$ and $s_{pq}$, and $k_{x,p}^2+k_{z,q}^2$.
\end{itemize}
Time-stepping:
\begin{itemize}
  \item Compute convolutions of $\tilde{u}_{x,pq}$, $\tilde{u}_{z,pq}$, $\tilde{\theta}_{pq}$ by FFT with 2/3 rule.
  \item Compute r.h.s. of evolution equations as
  
\begin{eqnarray*}
(\partial_t \tilde{u}_{x})_{pq} \leftarrow&
 ik_{x,p}(\tilde{u}_{x}*\tilde{u}_{x})_{pq}
+ik_{z,q}(\tilde{u}_{z}*\tilde{u}_{x})_{pq}
\\&-\sigma k_{pq}^2\tilde{u}_{x,pq} \\
(\partial_t \tilde{u}_{z})_{pq} \leftarrow&
 ik_{x,p}(\tilde{u}_{x}*\tilde{u}_{z})_{pq}
+ik_{z,q}(\tilde{u}_{z}*\tilde{u}_{z})_{pq}
\\&-\sigma k_{pq}^2\tilde{u}_{z,pq} + \sigma\tilde{\theta}_{pq} \\
(\partial_t \tilde{\theta})_{pq} \leftarrow&
 ik_{x,p}(\tilde{u}_{x}*\tilde{\theta})_{pq}
+ik_{z,q}(\tilde{u}_{z}*\tilde{\theta})_{pq}
\\&-k_{pq}^2\tilde{\theta}_{pq} + R\tilde{u}_{z,pq}.
\end{eqnarray*}
  \item Compute $(\partial_t\tilde{\mathbf u})$ as
  
\begin{eqnarray*}
(\partial_t \tilde{u})_{pq} \leftarrow&
\frac{-k_{z,q}(\partial_t\tilde{u}_x)_{pq}
      +k_{x,p}(\partial_t\tilde{u}_z)_{pq}}{k_{pq}^2} \\
(\partial_t \tilde{\mathbf u})_{pq} \leftarrow&
(-k_{z,q},k_{x,p}) \times \\
&\left[(\partial_t \tilde{u})_{pq}
     -\tilde{c}_{pq}\Sigma_{q'}(\partial_t \tilde{u})_{pq'}\right.\\
     &\left. -\tilde{s}_{pq}\Sigma_{q'}k_{z,q'}(\partial_t \tilde{u})_{pq'}\right].
\end{eqnarray*}
\end{itemize}
Here, we have denoted by $(\tilde{f}*\tilde{g})_{pq}$ the convolution product of fields $\tilde{f}_{pq}$ and $\tilde{g}_{pq}$ discretized on the wavenumber grid, which is performed by FFT.

It must be noted that the spatial discretization of the evolution equations has been performed in a closed form, independent of the time-stepping algorithm to be employed to solve the resulting set of ordinary differential equations, outlined above. We must also note that this system does not involve multiplication by matrices with dimension ${\cal N}$, the only nonlocal parts being the convolution products (handled through FFT) and the elimination of the longitudinal component of the velocity.

The time-stepping can then be performed by any algorithm designed to solve systems of ordinary differential equations. In our case, we opted for an adaptive-stepsize fifth order Runge--Kutta--Cash--Karp algorithm \cite{Press1996a}, which in our previous experience we have found efficient, stable, and flexible.

\section{Test runs}

Even for a simple system like the one we are studying here, the phenomenology found is rich; we present only a brief outline. All results are given in terms of the laterally-infinite cell crytical Rayleigh number $R_c\sim1701$, and the characteristic vertical diffusion time $t_c$ which for our cell and medium is $\sim11797$s. Coordinates and fields are in the dimensionless variables of Eqs. (\ref{eq1})--(\ref{eq3}). 

\begin{figure}[t]
\begin{center}
\includegraphics[width=1\columnwidth]{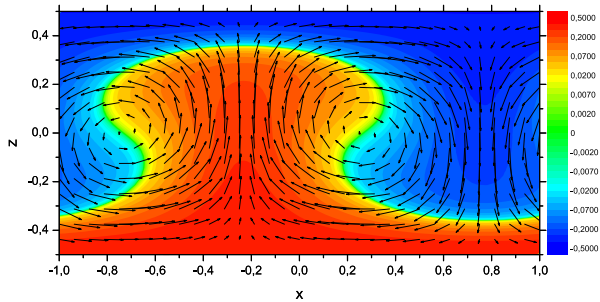}
\end{center}
\caption{Temperature and velocity fields for $R=5R_c$ at $t=2t_c$ on a 32$\times$16 grid.}
\label{fig1}
\end{figure}

\begin{figure}[t]
\begin{center}
\includegraphics[width=1\columnwidth]{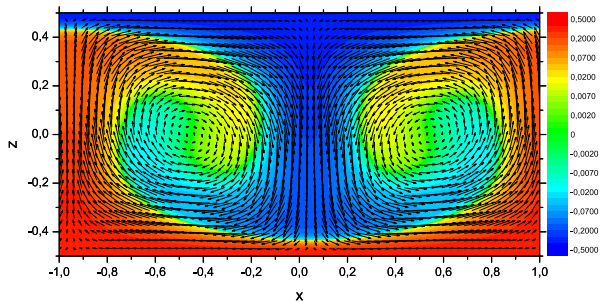}
\end{center}
\caption{Temperature and velocity fields for $R=50R_c$ at $t=t_c$ on a 64$\times$32 grid.}
\label{fig2}
\end{figure}

\begin{figure}[t]
\begin{center}
\includegraphics[width=1\columnwidth]{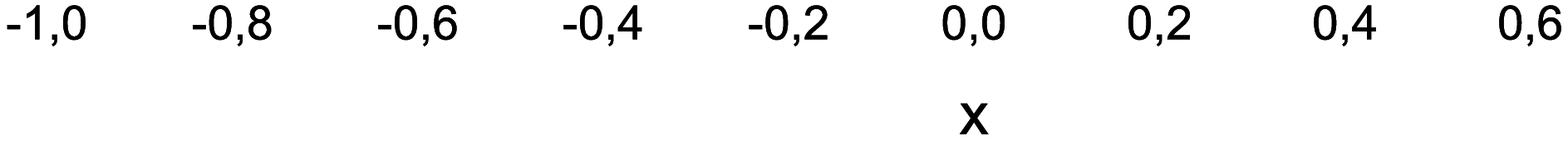}
\end{center}
\caption{Temperature and velocity fields for $R=500R_c$ at $t=0.25t_c$ on a 128$\times$64 grid (velocity decimated to a 64$\times$32 grid).}
\label{fig3}
\end{figure}

At low $R$, a stationary regime state, consisting in two counter-rotating rolls, is reached in times $\sim t_c$ or less. This time falls rapidly with increasing $R$, to $\sim 0.1t_c$ at $R\sim 1000R_c$ (see Figs.\ \ref{fig1}, \ref{fig2} and \ref{fig3}).

\begin{figure}[t]
\begin{center}
\includegraphics[width=1\columnwidth]{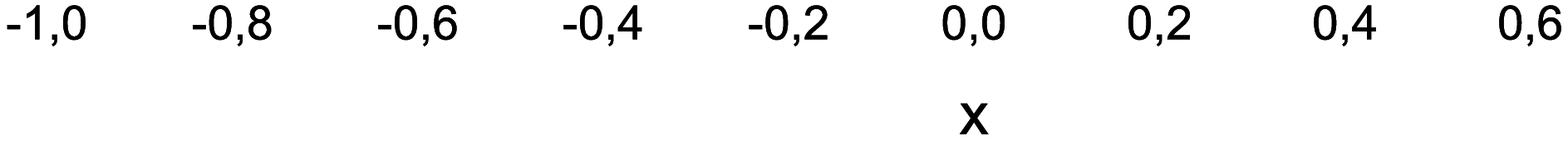}
\end{center}
\caption{Temperature and velocity fields for $R=5000R_c$ at $t=0.05t_c$ on a 192$\times$96 grid (velocity decimated to a 64$\times$32 grid).}
\label{fig4}
\end{figure}

Around $R\sim 5000R_c$, these rolls develop lateral oscillations, and the first ``secondary structures'' (small whirlpools) appear near the base of the ascending and descending plumes (see Fig.\ \ref{fig4}).

\begin{figure}[t]
\begin{center}
\includegraphics[width=1\columnwidth]{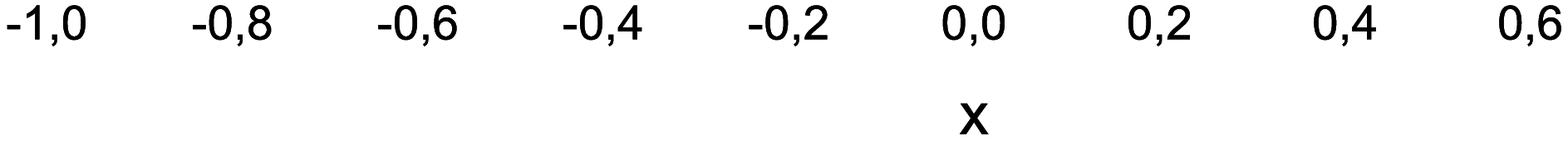}
\end{center}
\caption{Temperature and velocity fields for $R=5\times 10^4R_c$ at $t=0.01t_c$ on a 384$\times$192 grid (velocity decimated to a 64$\times$32 grid).}
\label{fig5}
\end{figure}

Above $R\sim 5000R_c$, the regime state becomes disordered and aperiodic, consisting of intermittent plumes and whirlpools in a wide size range (see Fig.\ \ref{fig5}).

At $R\sim 5\times10^5R_c$, the temperature difference is $\sim 65$K; the smallest whirlpools are $\sim 1$cm wide, and the typical wind speeds are $\sim 1$m/s, in agreement with experimental observations \cite{DePaul2009} (see Fig.\ \ref{fig6}). The time to reach this regime state is rather short, $\sim 0.001t_c$.

Note that in all figures, except for Figs.\ \ref{fig1} and \ref{fig2}, the velocity grid has been decimated to enhance clarity.

\section{Code performance}

Over the full range of $R$ tested here (more than five orders of magnitude), the code maintained the typical velocity divergence and the field values at the bottom and top boundaries at essentially machine-precision zero, showing that the implemented method is sound.

Also over this range of $R$, the grid spacing needed to achieve ``smooth'' fields ({\it i.e.}, to capture all the physical detail down to the smallest present scales) is consistent with the width of the (thermal) boundary layer. However, for coarser grids the code still gives qualitatively sound results; typically a checkerboard-like instability develops, but the algorithm keeps it quenched, showing very good stability even in presence of a severe accuracy loss.

\begin{figure}[t]
\begin{center}
\includegraphics[width=1\columnwidth]{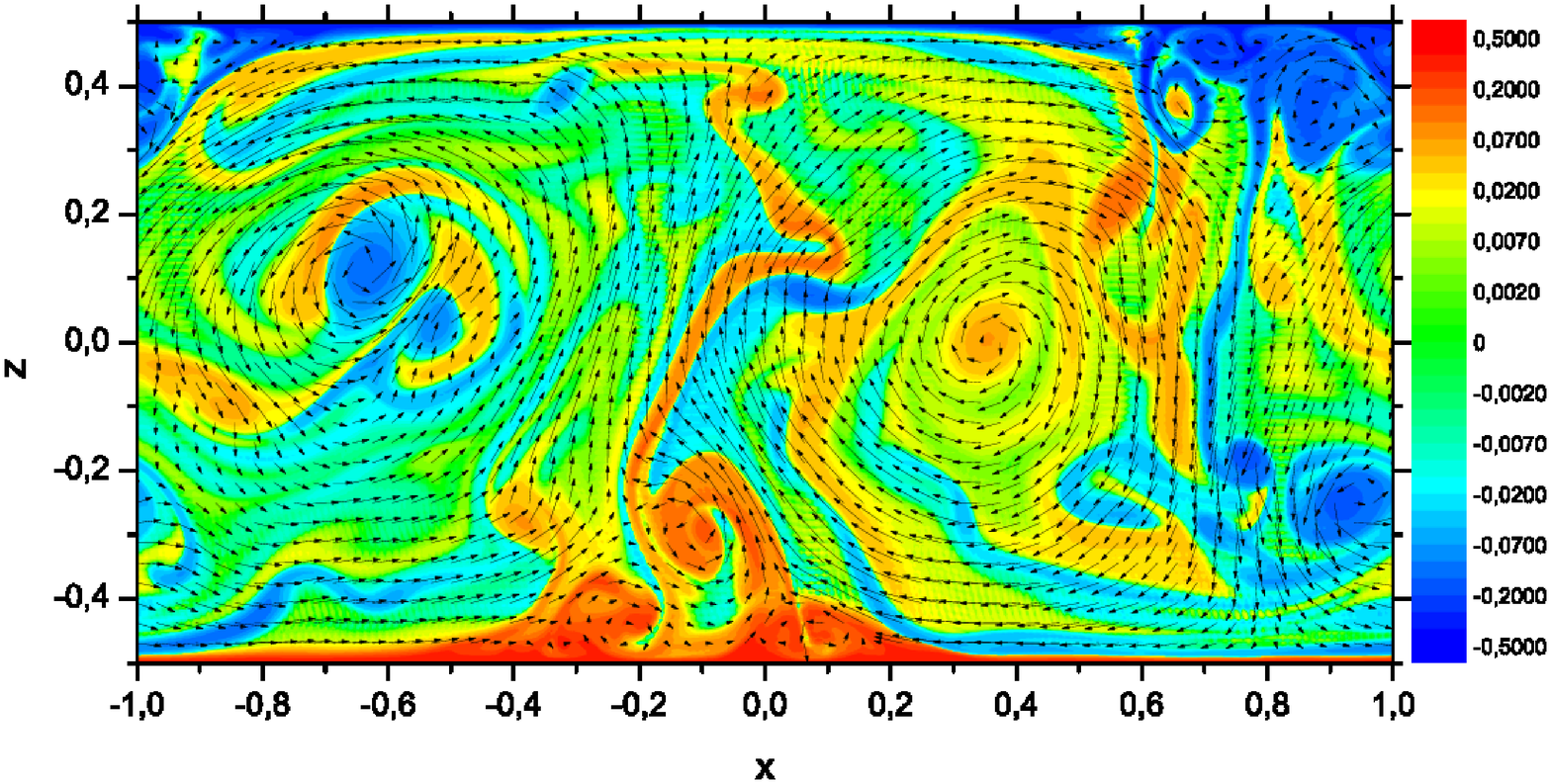}
\end{center}
\caption{Temperature and velocity fields for $R=5\times 10^5R_c$ at $t=0.002t_c$ on a 512$\times$256 grid (velocity decimated to a 64$\times$32 grid).}
\label{fig6}
\end{figure}

The algorithm is also fast: the simulation for $R\sim 10^9$ (see Fig.\ \ref{fig6}), on a 512$\times$256 grid, took less than one day per simulated minute on a single core of the 3GHz PentiumD CPU on which all our runs were performed, with no code optimization. 

It is difficult to find in the literature a directly comparable simulation for more accurate benchmarking. However, from Ref. \cite{Mercader2009} we can see that, for example, the relaxation to a (steady) regime state in Rayleigh--B\'enard convection at $R\lesssim 30000\sim 20 R_c$ on a $\sim 20000$-node grid takes 46 minutes on a similar processor at similar speed (3.2GHz Pentium 4), by the method implemented there. In the case of our code, at $R=20 R_c$ on a $200\times 100$ grid, and with the sole optimization of grouping the real FFTs in pairs (see the 2FFT algorithm in Ref. \cite{Press1996a}), the equivalent relaxation took 8 minutes, which is shorter by a factor of $\sim 6$. But it must be taken into account that with our initial conditions (zero velocities and thermal noise in temperatures) the convection onset is slow, and is followed by a transient stage with strong and disordered convective patterns that decay to the regime state very slowly.

\section{Conclusions and outlook}

We have been able to show that a Fourier-based pseudospectral method can be adapted to a (admittedly simple) non-free BC setting, at the cost of moderate analytical work on the solutions of Poisson's and Laplace's equations. The method is formulated in primitive variables, but the pressure is not explicitly computed nor referenced, like in a streamfunction-vorticity formulation. It also shares some properties with projection methods, but it decouples the implementation of the incompressibility condition from the time-stepping scheme, allowing great flexibility in the selection of the last. The resulting code is fast and stable, and implements the BCs and the incompressibility condition essentially to machine-precision.

Work on the extension of this scheme to a fully closed Rayleigh-B\'enard cell ({\it i.e.}, with non-free BCs also on the lateral walls) is currently under course.

\end{document}